\documentclass[traditabstract]{aa} 
\usepackage{epsfig}
\usepackage{graphicx}
\usepackage{upgreek}
\usepackage{xcolor}
\usepackage{txfonts}
\usepackage{amssymb}
\usepackage{natbib}                          
\newcommand{\be}{\begin{equation}}
\newcommand{\ee}{\end{equation}}
\newcommand{\bea}{\begin{eqnarray}}
\newcommand{\eea}{\end{eqnarray}}

\begin{document}

\title{Probing the possibility of hotspots on the central neutron star in HESS\,J1731$-$347} 
\author{
V.~F. Suleimanov\inst{1,2},
D. Klochkov\inst{1},
J. Poutanen\inst{3,4},
\and 
K. Werner\inst{1}}


\institute{
{Institut f\"ur Astronomie und Astrophysik, Kepler Center for Astro and
Particle Physics, Universit\"at T\"ubingen, Sand 1,
 72076 T\"ubingen, Germany  \email{suleimanov@astro.uni-tuebingen.de}}
\and
Space Research Institute of the Russian Academy of Sciences, Profsoyuznaya Str. 84/32, Moscow 117997, Russia
\and
{Tuorla Observatory, Department of Physics and Astronomy, University of Turku, V\"ais\"al\"antie 20, FI-21500 Piikki\"o, Finland}
\and
{Nordita, KTH Royal Institute of Technology and Stockholm University, Roslagstullsbacken 23, SE-10691 Stockholm, Sweden}
}

\date{Received xxx / Accepted xxx}

   \authorrunning{Suleimanov et al.}
   \titlerunning{Hotspot probing for CCO in HESS\,J1731}

\abstract
{
The X-ray spectra of the neutron stars located in the centers of supernova remnants Cas A 
and HESS\,J1731$-$347 are well fit with carbon atmosphere models. 
These fits  yield  plausible neutron star sizes for the known or estimated distances to these supernova remnants. 
The evidence in favor of the presence of a pure carbon envelope at the neutron star surface 
is rather indirect and is based on the assumption
 that the emission is generated uniformly by the entire stellar surface. 
Although this assumption is supported by the absence of pulsations, the observational upper
 limit on the pulsed fraction is not very stringent.
In an attempt to quantify this evidence, we investigate the possibility that the observed spectrum 
of the neutron star in HESS\,J1731$-$347
 is a combination of the spectra produced in a hydrogen atmosphere  of the hotspots and of 
 the cooler remaining part of the  neutron star surface. 
The lack of pulsations in this case has to be explained either
by a sufficiently small angle between the neutron star spin axis and the
 line of sight, or by a sufficiently small angular distance between the hotspots and the neutron star rotation poles.
As the observed flux from a non-uniformly emitting neutron star depends on the angular distribution of the radiation 
emerging from the atmosphere, we have computed two new grids of pure carbon and pure 
hydrogen atmosphere model spectra accounting for Compton scattering.  
Using new hydrogen models, we have evaluated the probability of a geometry that leads 
to a pulsed fraction below the observed upper limit to be about 8.2\%. 
Such a geometry thus seems to be rather improbable but cannot be excluded at this stage.
}

\keywords{ radiative transfer  --  methods: numerical  -- stars: atmospheres -- stars: neutron -- X-rays:stars}

\maketitle
%

\section{Introduction}

Measurements of neutron star (NS) radii are of fundamental importance not only for the astrophysics of these objects.
NS radii reflect the properties of the cold supra-dense matter in their cores, such as the rigidity, and constrain the equation of state
 (EoS) of cold dense matter (see reviews by \citealt{LP16} and \citealt{ML16}). 
A variety of approaches has been undertaken to evaluate the NS radii (see the aforementioned reviews). 
Here, we consider the approach based on modeling the NS thermal X-ray spectrum. 

The sizes of some NSs can be estimated from a fit of their X-ray spectrum with a model function.
In the first approximation, spectra of thermally emitting cooling NSs have blackbody-like shapes.
However, the blackbody fit underestimates the NS size,  giving, for example, a radius below 
1 km for the NS in the center of supernova remnant Cas~A \citep{Pavlov.etal:00}.
This object belongs to a subclass of NSs in supernova remnants, which do not demonstrate any radio-pulsar 
or magnetar activity and are dubbed central compact objects \cite[CCOs, see, e.g.,][]{Gotthelf.etal:13}. 
We mainly consider these NSs in this work. 

The main reason why the blackbody fits underestimate the NS radii is that the emerging thermal radiation
  forms in the gaseous atmosphere of the star.  
Photons emerging from the atmosphere prefer to escape  at highest possible energies because the plasma opacity 
decreases toward higher  energies and the temperature increases with the depth in the atmosphere 
\citep[see details in the reviews by][]{Zavlin:09, Potekhin:14,S.etal:16}.  
As a result, the thermal atmosphere  spectrum is harder than the blackbody spectrum with the same effective temperature. 
It has, however, a significantly lower flux spectral density than a blackbody spectrum with the same color temperature. 
Therefore, the atmosphere models require  a larger emission area than the blackbody fit, which is 
more consistent with the expected NS radii. 

Gravitational separation of elements is very effective in the NS envelopes \citep{BSW92}. 
Therefore it is very reasonable to assume that NS atmospheres are chemically pure and consist of the lightest element.   
However, the hydrogen  atmosphere model fit to the spectrum of the CCO in Cas~A still 
yields  an unrealistically small NS radius of $\sim$5\,km \citep{PL09}. 
Only the fit with carbon atmosphere model spectra gives an acceptable NS radius \citep{HH09}. 
Recently, the X-ray spectrum of several other CCOs have been successfully fitted with carbon model atmospheres  
\citep{Klochkov.etal:13, Klochkov.etal:16}, suggesting that a significant fraction of young cooling 
NSs may have carbon envelopes. 

On the other hand, there is no direct evidence that carbon dominates the atmospheres of the considered NSs.
Hydrogen atmospheres cannot be excluded when we assume that the NS surface is inhomogeneous 
and most of the radiation is emitted from hot spots. 
In particular, \citet{Bogdanov:14} showed that the phase-averaged X-ray spectrum of the CCO in the supernova remnant
 Kes~79 can be well fit with a carbon atmosphere spectrum and give a realistic NS radius. 
Strong pulsations with the pulsed fraction ($PF= (F_{\rm max}-F_{\rm min})/(F_{\rm max}+F_{\rm min})$, 
where $F_{\rm max}$ and  $F_{\rm min}$ are the 
maximum and the minimum fluxes) of $\sim$64\% are, 
however, observed from this CCO \citep{HG10}, clearly showing that the emission is not uniformly distributed across the stellar surface.

X-ray pulsations have not been detected from the CCOs in Cas~A and HESS\,J1731$-$347 so far, but the  upper limits on the 
$PF$ are not very stringent \citep[$\sim$7--8\% and 12\%, see][]{Klochkov.etal:15, HG:10}. 
The possibility of an inhomogeneous hydrogen-emitting stellar surface (hotspots) can therefore not 
be completely excluded and has to be investigated.  
The fit with hydrogen atmospheres gives an effective temperature higher than the fit with  carbon atmospheres 
($\sim$3\,MK vs. $\sim$2\,MK). 
The temperature of hotspots can be even higher. 
In this case, Compton scattering has to be taken into account because it changes the hard tails of 
the hydrogen atmosphere model spectra  \citep{SW07}. 
If  the NS emits non-uniformly, the observed flux depends not only on the flux escaping from the atmosphere, 
but also on the angular distribution of the radiation. 
This effect is most pronounced for  fast-rotating NSs, 
 as was illustrated in the modeling of the X-ray light curves from millisecond-accreting and radio pulsars aimed 
 at the measuring of their radii \citep{PG03,Bogdanov:16}.

In this paper, we reanalyze the X-ray spectrum of the brightest known CCO, the NS in the supernova remnant
 HESS\,J1731$-$347 \citep[see][for details]{Klochkov.etal:15}. 
For this study, we have computed two new extended model grids of NS
 atmospheres consisting of pure hydrogen and pure carbon. These grids also 
account for Compton scattering and include information on the angular distribution of 
radiation that escapes from the atmosphere. 
We first fit the X-ray spectrum of HESS\,J1731$-$347 with the carbon atmosphere model spectra, which shows that the Compton effect does
 not significantly affect the constraints  on the NS mass and radius. 
We then consider the possibility that the emission from this NS is produced at the non-uniformly 
emitting  hydrogen atmosphere of two different temperatures. 
This two-component model provides as good a fit to the data as the carbon models. 
Assuming that the hotter region is associated with the magnetic poles, we finally evaluate the probability of a geometry 
that is consistent with the observed upper limits on the $PF$.
A similar analysis for NSs in quiescent low-mass X-ay binaries was performed by \citet{EHMBS16}.

\section{Atmosphere models}

The input parameters of model atmospheres are the effective temperature, $T_{\rm eff}$, the surface gravity, 
$g$,  and the chemical composition.  
The basic equations determining the model atmosphere structure and the spectrum that emerges from the model when Compton scattering is accounted for, as well as the methods of their solution  were presented in our earlier works
 \citep{SPW11, SPW12}. 
The computation details of the carbon atmosphere models used in the current work can be found in \citet{S.etal:14}. 

Here we consider NS atmospheres that only consist of pure hydrogen and pure carbon. 
For each of these two compositions, we computed a grid of models for nine values of $\log g$, from 13.7 to 14.9 with a step of 0.15.  
We considered the effective temperature to range from 0.5 to 10\,MK with a step of 0.05 MK for pure hydrogen
 (191 models for each $\log g$) and from 1.0 to 4.0\,MK  with the same step for pure carbon (61 models for each $\log g$). 
 
\begin{figure}
\centering
\includegraphics[angle=0,width=\columnwidth]{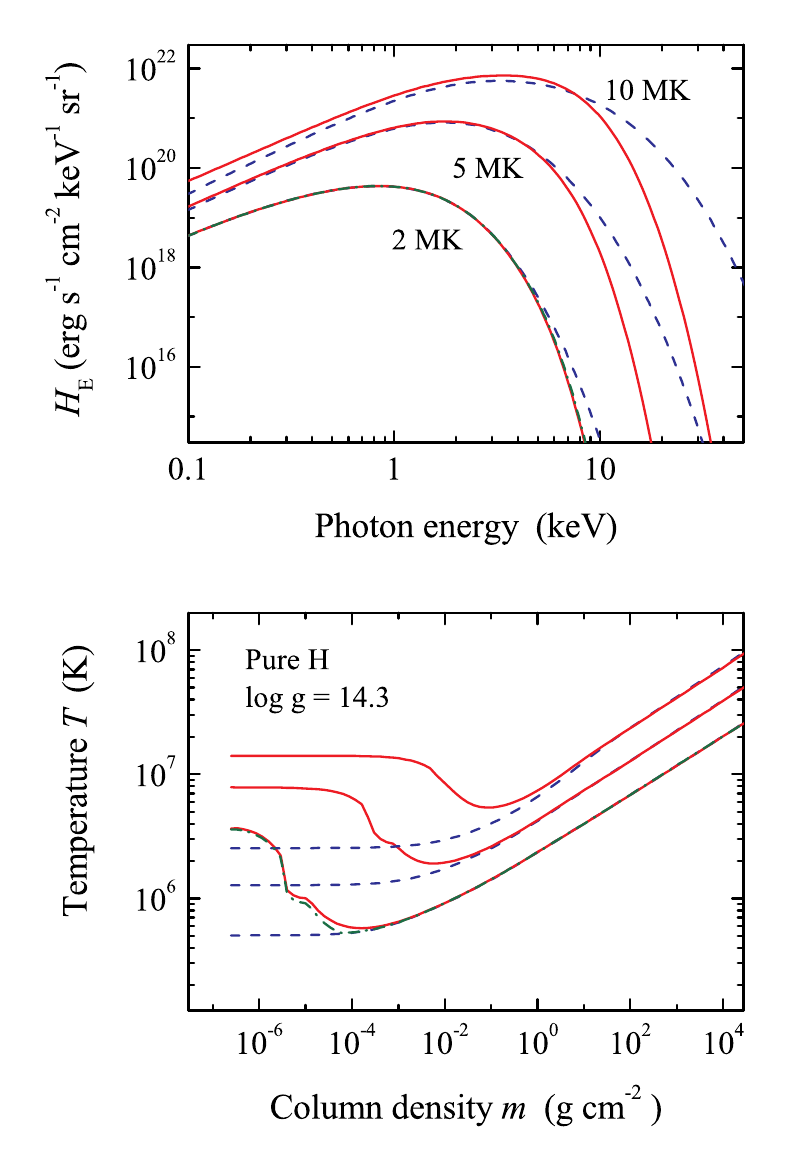}
\caption{\label{fig1}
Emerging intrinsic model spectra ($4\pi H_E$ is the radiation flux) of pure hydrogen NS 
atmospheres (top panel) computed with (red solid curves) and without (blue dashed curves) Compton
scattering. 
The temperature structures of the same models are shown in the bottom
panel. The spectrum of the model computed using the Kompaneets
equation and its temperature structure are shown with green dot-dashed curves.
}
\end{figure}
 
\begin{figure}
\centering
\includegraphics[angle=0,width=\columnwidth]{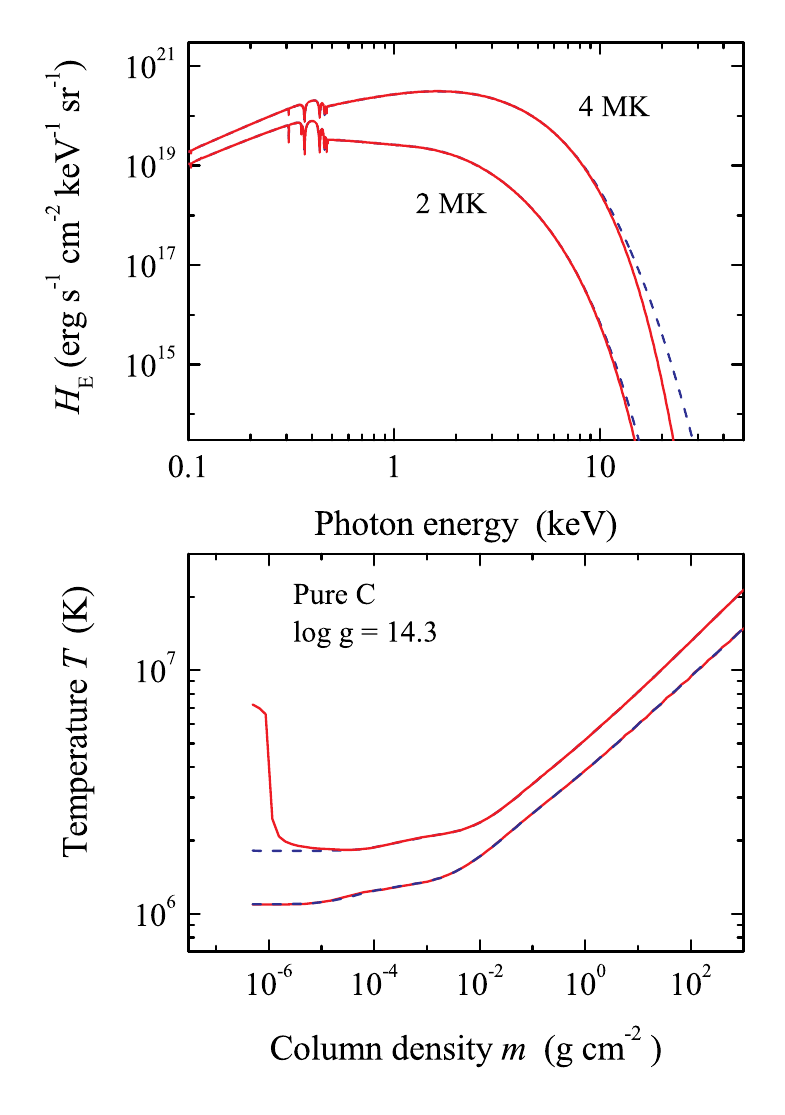}
\caption{\label{fig2}
Same as in Fig.\,\ref{fig1}, but for  pure carbon model atmospheres.
}
\end{figure}

\subsection{Computation methods and general properties of the atmospheres}

The new sets of NS model atmospheres and their emerging spectra with Compton scattering were computed using two methods. 
The first method is described in \citet{SPW12}, where the exact angle-dependent redistribution function is used. 
We used this approach only for high-temperature pure hydrogen models with $T_{\rm eff}>$\,2\,MK.  
The hydrogen models with lower temperatures and all the carbon models were computed using the Kompaneets
equation approximation for  Compton scattering \citep{SPW11}. 

Absorption lines are a significant opacity source in carbon atmospheres. 
We therefore need a dense frequency grid to describe
them correctly. 
Unfortunately, the current computation facilities do not permit usaging more than 
$\sim$1000 frequency points in the existing codes \citep{SPW11, SPW12}. 
To overcome this difficulty, we divided the photon energy band into two parts in which the electron scattering is treated differently.
The electron scattering  at the low-energy part ($E < 0.5$\,keV), where the absorption lines are located and a large number of 
 frequency points is necessary, is  approximated as coherent Thomson scattering. 
Compton scattering at higher energies is described using the Kompaneets equation \citep{SPW11}, which accounts for photon
 redistribution between neighboring photon energy bins.  
The high-energy band is taken above the \ion{C}{VI} photoionization edge (0.49 keV) such that 360 frequency points are 
sufficient for describing a smooth continuum in this band. 
However, the mean intensity $J_\nu$ is summed over the whole  energy band in the energy 
balance equation, which determines a model temperature structure 
\begin{equation}  
\label{econs}
\int_0^{\infty}\! \! \! \! \!  \kappa_{\nu}\left(J_{\nu} - B_{\nu}\right) {\rm d}\nu =   \kappa_{\rm e} \frac{kT}{m_{\rm e} c^2} 
 \left[ 4\! \!  \int_0^{\infty} \! \! \! \! \! \! J_{\nu} {\rm d}\nu -  \! \!  \int_0^{\infty}\! \! \!\! \!\! \!    x J_{\nu}
\left( 1+\frac{CJ_{\nu}}{x^3}\right)  {\rm d} \nu \right].
\end{equation}    
Here $x=h \nu /kT$ is the dimensionless frequency,
$B_{\nu}$ is the blackbody (Planck) 
intensity, $T$ is the local electron temperature, $\kappa_{\nu}$ is the
opacity per unit mass due to free-free, bound-free, and bound-bound
transitions, $\kappa_{\rm e}$ is the electron (Thomson) scattering
opacity, and $C=c^2 h^2~/~2(kT)^3$. 

The same approach is used for the computations of the low-temperature hydrogen models. 
In this case, the electron scattering is treated as coherent at photon energies below 0.1 keV. 
The boundary model, with $T_{\rm eff}$\,= 2\,MK, is computed using both approaches to treat Compton scattering and 
 gives very similar results (see Fig.\,\ref{fig1}).  

Some examples of the model spectra computed with and without Compton scattering are
 shown in Figs.\,\ref{fig1} and \ref{fig2}. 
As expected \cite[see, e.g.,][] {S.etal:16},  if Compton scattering is accounted for, the high-energy tails of the emerging spectra become close to diluted blackbodies, and chromosphere-like surface layers with increased temperature arise. 
We note that the upper chromosphere-like layer is heated by hard photosphere photons as a result of the Compton effect,
 and no additional external heating (for instance,  low-rate accretion) is necessary. However, low-rate accretion can make a
chromosphere-like surface more prominent \citep[see, e.g., ][]{ZTZT95, ZTT98, ZTT00}. Therefore, 
correct consideration of the Compton scattering will increase the model flux at the UV  and optical spectral bands by a few
percent. This possibility has to be taken into account when investigating spectral properties of isolated NSs.

Compton scattering is more important for hydrogen than carbon atmospheres because the electron opacity 
depends on the hydrogen mass fraction $X$ as $\kappa_{\rm e} \approx 0.2(1+X)$\,cm$^2$\,g$^{-1}$.  
Furthermore, the true opacity (bound-free and free-free absorption opacity $\kappa_{\nu}$)  is stronger in the carbon atmospheres 
 because carbon is only partially ionized and because of a $Z^2$ factor in free-free absorption.

\subsection{Angular distribution of emergent radiation}

\begin{figure}
\centering
\includegraphics[angle=0,width=\columnwidth]{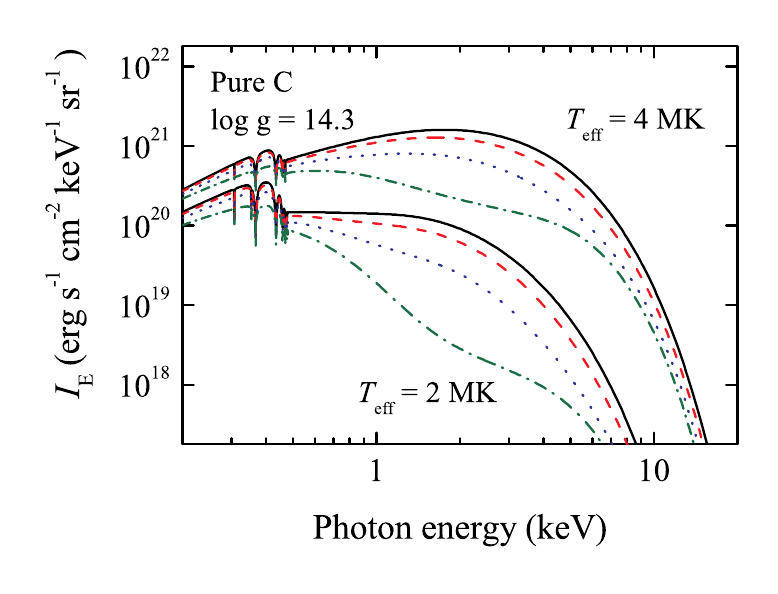}
\caption{\label{fig3}
Model spectral energy distributions of specific intensities for
the carbon atmospheres with $\log g =14.3$ and two $T_{\rm
  eff}$\,= 2 and 4\,MK, 
computed for four angles, $\mu=\cos \theta$\,= 0.9306 (solid curves),
0.67 (dashed curves),  
0.33 (dotted curves), and 0.0694 (dash-dotted curves).
}
\end{figure}

\begin{figure}
\centering
\includegraphics[angle=0,width=\columnwidth]{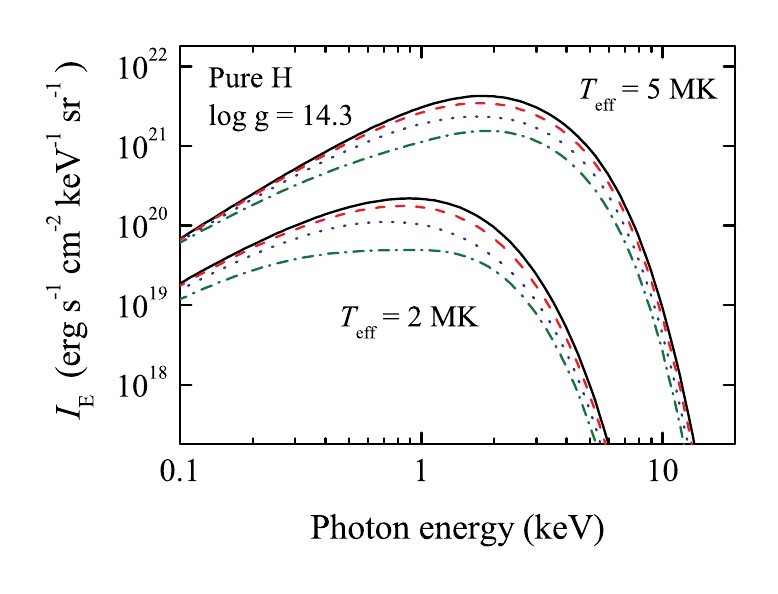}
\caption{\label{fig4}
Model spectral energy distributions of specific intensities for
the pure hydrogen atmospheres with $\log g =14.3$ and two
$T_{\rm eff}$\,= 2 and 5\,MK, 
computed for the same four angles as in Fig.\,\ref{fig3}.
}
\end{figure}

As described above, our model grid contains information about the angular distribution of 
the emerging radiation in addition to the flux spectra. 
For each model, we computed spectral distributions of the specific intensities in four angles chosen  as the nodes of the 
Gaussian quadrature in the [0,1] interval, with cosines of the angles to the normal $\mu = \cos \theta$ 
being equal to approximately  0.9306, 0.6700, 0.3300, and 0.0694.  
The specific intensity spectra of some of our models are presented in Figs.\,\ref{fig3} and \ref{fig4}. 
 
The main feature of the angular dependence in the carbon model intensity spectra is the fact that at photon energies above 1 keV,
the emerging radiation is strongly peaked along the normal to the stellar surface, and the specific intensity  at the highest angle 
to the normal is almost two orders of magnitude lower than the intensity corresponding to the normal. 
As is well known \citep{RL:79}, the emerging intensity at angle $\arccos\mu$ approximately corresponds to the
 Planck function at the  thermalization depth $m^*_{\nu}$ from which the emerging  photons escape: 
\be
     \tau^*_{\nu} =\int_0^{\,m^*_\nu}\,\frac{1}{\mu}\sqrt{\kappa_{\nu}(\kappa_{\nu}+\kappa_{\rm e})}\, {\rm d}m \approx 1. 
\ee  
The considered carbon atmospheres have a significantly higher true absorption opacity at the low-temperature layers  on the
surface because carbon becomes almost fully ionized in the deep high-temperature layers.  
Therefore, the geometrical path along a ray that is highly inclined to the normal is much shorter than the path along the normal. 
The escaping photons are thus created in a layer of a substantially lower temperature in this case, 
resulting in a strong beaming of the emerging radiation along the normal. 

The feature of the carbon atmospheres described above occurs
when the true opacity $\kappa_{\nu}$ is
 comparable to the electron scattering opacity. 
Hence, the hydrogen model atmospheres only show emerging radiation that peaks along the normal for the low-temperature atmospheres
because the maximum of the spectral flux shifts to lower photon energies where the free-free opacity is higher. 
The decrease in temperature is also important because the free-free opacity depends on temperature as well: 
\be
     \kappa_{\nu, \rm ff} \sim \rho\, T^{-1/2}\, \nu^{-3}\, Z^2.
\ee
However, these effects are less significant than the carbon ionization. 
As a result, the  intensity of radiation that emerges from hydrogen atmospheres  is not so strongly peaked along the normal. 
Electron scattering dominates in high-temperature hydrogen atmospheres ($T_{\rm eff} \ge $\,3 -- 5\,MK) where the angular 
distribution of the emerging radiation is close to the well-known
distribution for an atmosphere that is dominated by electron
scattering \citep{ChaBre47,Cha60, sob49, sob63}.

\begin{figure}
\centering 
\includegraphics[angle=0,width=\columnwidth]{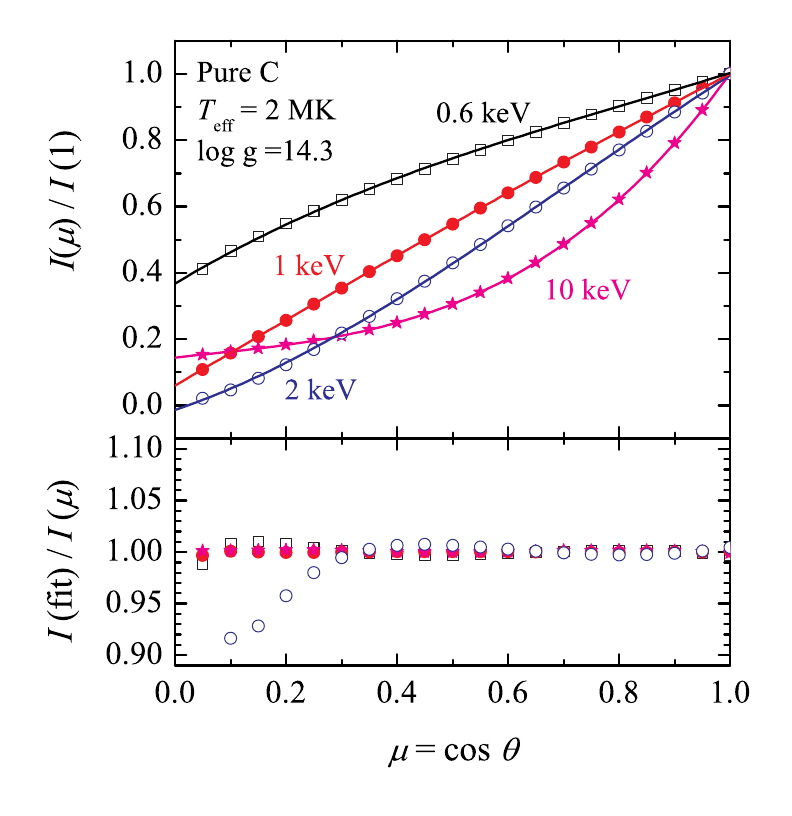}
\caption{\label{fig5}
Top panel:\textup{} comparison of the specific intensity angular distributions computed
for 20 angles (symbols) with the polynomial fits based on four angles (see Fig.\,\ref{fig3}). 
The results are presented for four photon energies and for the $T_{\rm eff}$\,=2\,MK carbon atmosphere model.
The relative accuracy of the fits is shown in the  {\it \textup{bottom panel}}. 
}
\end{figure}

\begin{figure}
\centering
\includegraphics[angle=0,width=\columnwidth]{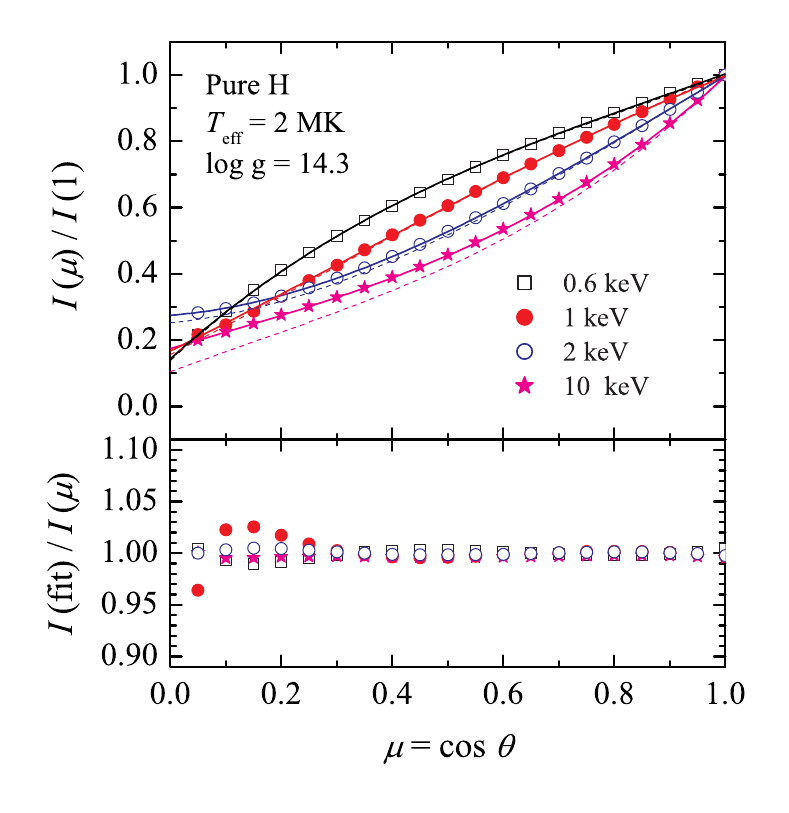}
\caption{\label{fig6}
Same as in Fig.\,\ref{fig5}, but for the hydrogen model atmospheres computed using Kompaneets equation. 
The dashed curves correspond to the models computed with the exact Compton redistribution function. 
}
\end{figure}

We can now fit the angular distribution of the emerging specific intensity $I_E(\mu)$ at every photon energy 
$E$ by cubic polynomials  
\be \label{mufit}
            I(\mu) = I(1)\, (a+b\,\mu+c\,\mu^2+d\,\mu^3), 
\ee  
where the coefficients $a$, $b$, $c$, and $d$ are straightforwardly defined by the known $I_E(\mu)$ in the four angles.  

We have checked the accuracy of this approximation for two model atmospheres and for four photon energies. 
We computed specific intensities for 20 angles equidistantly spaced in $\mu$  at these photon energies 
for the given model atmospheres
and compared the obtained values with those computed using the polynomial fits. 
The results are shown in Figs.\,\ref{fig5} and \ref{fig6}.
The accuracy of the cubic polynomial fits is better than 1\%  at almost all angles excluding 
the largest angles ($\mu < 0.2-0.3, \theta > 70\degr-80\degr$). 
Significant deviations at large angles are typical of the most normally peaked angular distributions, that is, at 2 keV for the carbon model
atmosphere and at 1 keV for the hydrogen model atmospheres. 
The less peaked distributions show less deviations; they remain in the range 1--2\% even at large angles. 
We therefore conclude that the cubic polynomial fits provide sufficiently accurate approximations
 to the angular distributions of the emerging intensities in
the model. 
The contribution to the total observed flux from those parts of the NS surface that are highly inclined 
with respect to the line of sight is
 low because of the small projected areas of these parts and because of the low specific intensities at these angles. 
 
We also compared the angular distributions computed for the hydrogen model atmosphere
 ($T_{\rm eff} =2$\,MK, $\log g = 14.3$) 
using two different approaches to describe Compton scattering: using the 
Kompaneets equation, and using the exact angle-dependent redistribution function. 
The fits obtained using the latter approach are shown with dashed curves in Fig.\,\ref{fig6}. 
The accurate redistribution function gives more strongly peaked 
angular distributions with the most 
apparent deviations at higher photon energies and larger angles. 
The reason is that  the Kompaneets equation approach assumes isotropic scattering.

\section{Results of the spectral modeling}
\label{sec:resspec}

\subsection{Fitting of the observed spectrum}

\begin{figure}
\centering
\includegraphics[angle=0,width=\columnwidth]{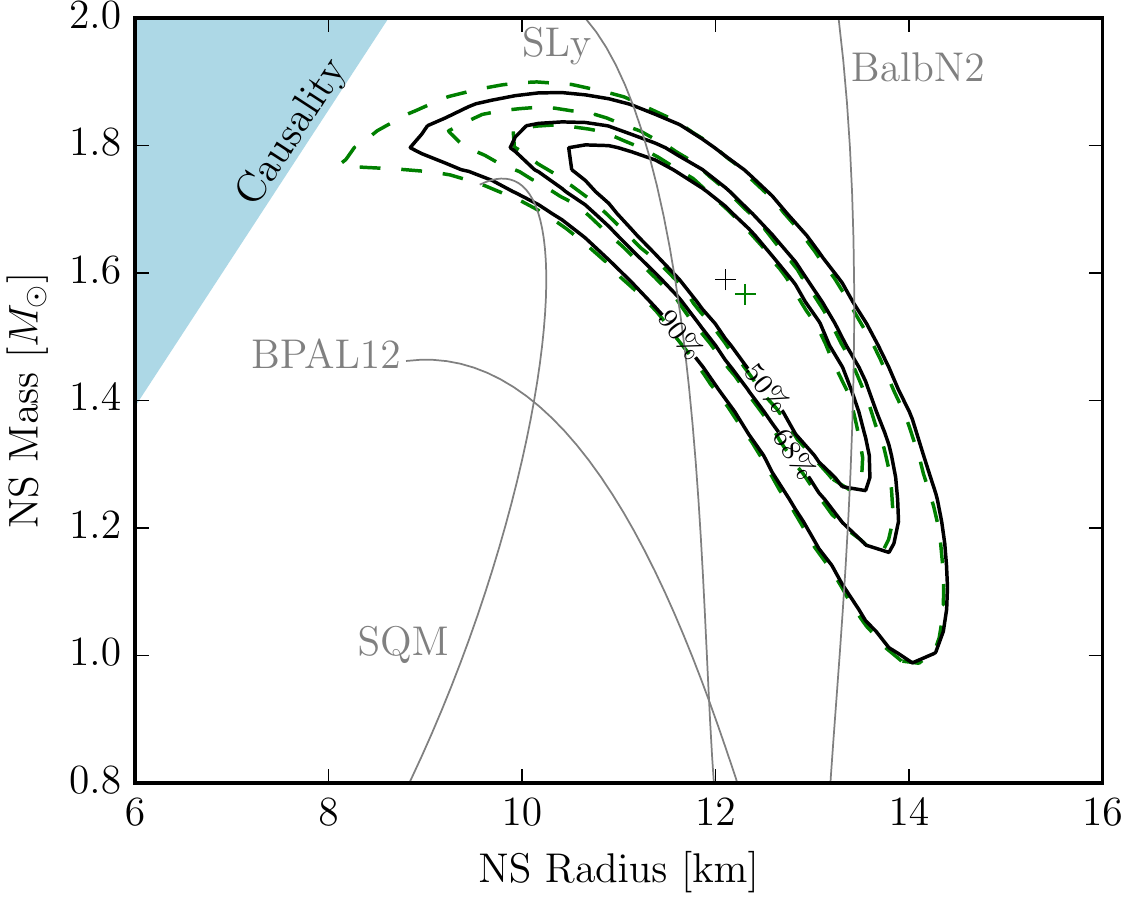}
\caption{\label{figRMC}
Comparison of the confidence regions on the mass--radius plane for the CCO in HESS\,J1731$-$347 using the spectra 
of the pure carbon atmosphere models  computed with (solid black contours) and without (dashed green contours)
Compton scattering. 
The contours correspond to 50, 68, and 90\% confidence levels for the two parameters of interest.
The assumed distance to the source is 3.2 kpc. The reduced $\chi^2_{\rm red}\approx 1.07$ (895 d.o.f.).
The thin gray curves indicate the mass--radius dependencies for some of the commonly used nuclear EoS.
}
\end{figure}

\begin{figure}
\centering
\includegraphics[angle=0,width=\columnwidth]{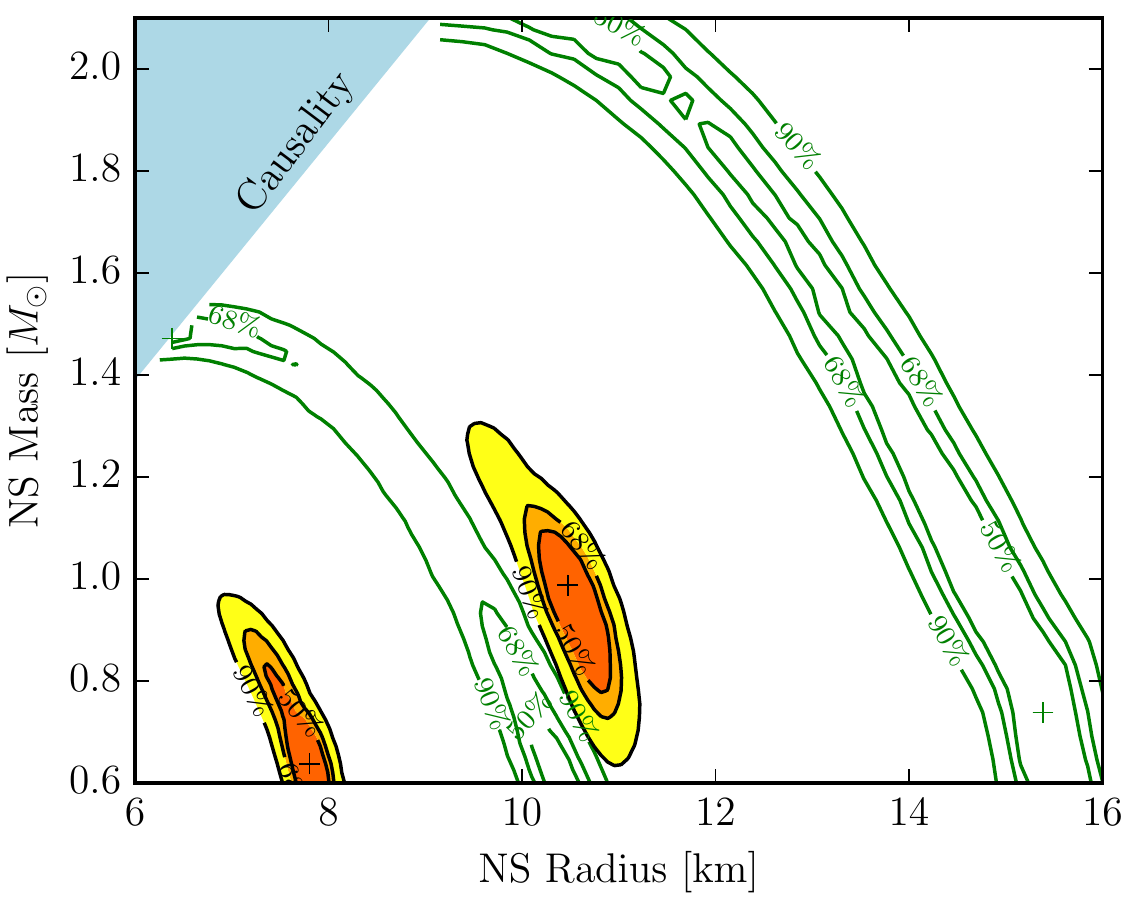}
\caption{\label{fig8}
Same as Fig.\,\ref{figRMC}, but for the pure hydrogen atmosphere models. 
The contours corresponding to the models that account for Compton scattering are colored. 
The assumed distances are  7 kpc (lower left contours) and 10 kpc (upper right contours).
}
\end{figure}
 
\begin{figure}
\centering
\includegraphics[angle=0,width=\columnwidth]{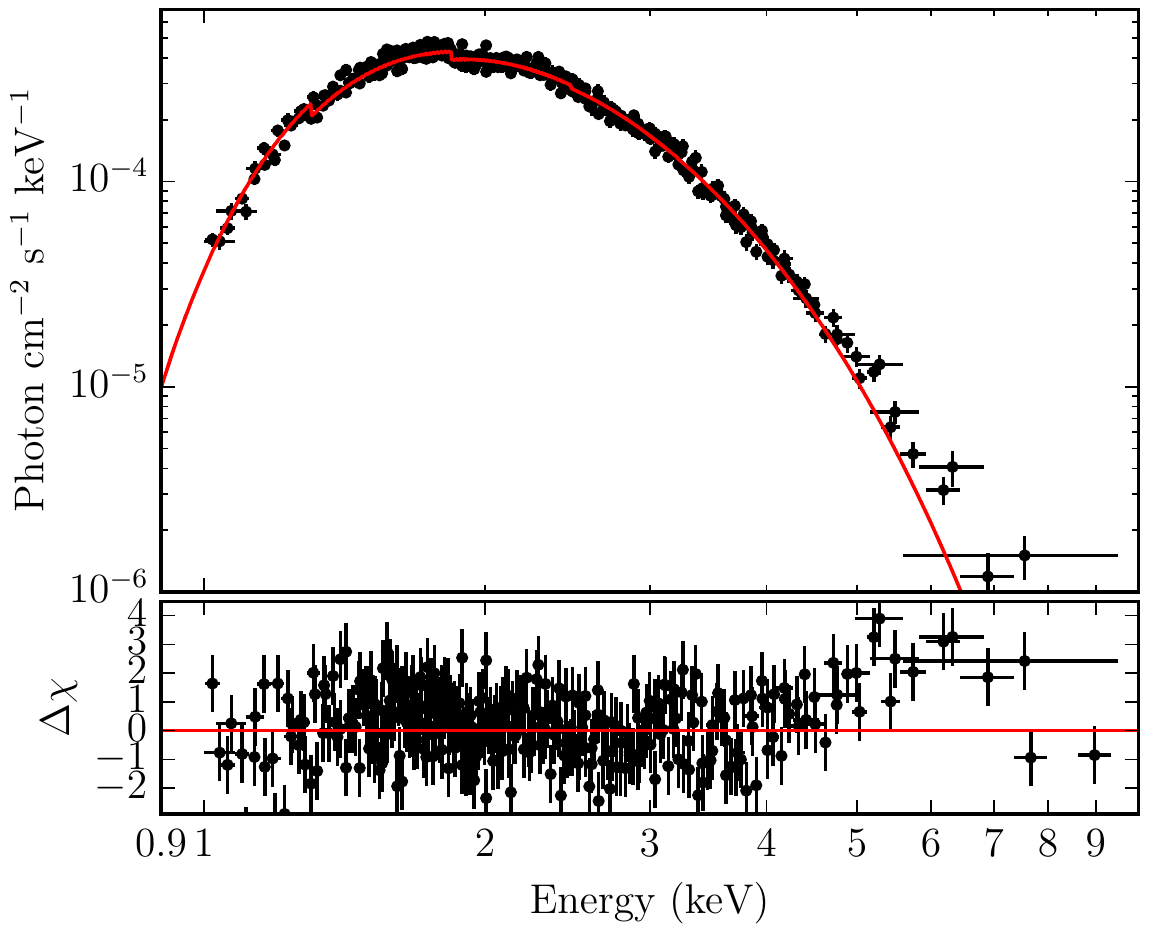}
\caption{\label{fig9}
Combined \emph{XMM-Newton} spectrum of the CCO in HESS\,J1731$-$347, 
published in \citet{Klochkov.etal:15}, fitted with our hydrogen atmosphere model
spectrum accounting for   Compton scattering for the fixed distance of 10 kpc (top panel) 
and the corresponding fit residuals (bottom panel\textup{}).
The reduced $\chi_{\rm red}^2 \approx$\,1.17 corresponding to the
null-hypothesis probability of $P$ = 0.0002, which indicates an unacceptable fit.
}
\end{figure}

We used the newly computed model tables, which we have also made available in {\sc xspec}
\footnote{http://heasarc.gsfc.nasa.gov/xanadu/xspec/models/hatm.html,
https://heasarc.gsfc.nasa.gov/xanadu/xspec/models/carbatm.html}, 
to fit the observed X-ray spectrum of the CCO in HESS\,J1731$-$347 presented in \citet{Klochkov.etal:15}.  
A comparison of the old and new confidence regions in the $M-R$ plane is shown in Fig.\,\ref{figRMC} for pure carbon atmosphere models.
The two fits have equal quality ($\chi^2_{\rm red} \approx 1.07$). 
As expected, the differences are minor and more obvious at the high-mass -- small-radius end of the confidence regions where the 
intrinsic effective temperatures of the models  are the highest and the Compton scattering contribution is largest.  
On the other hand, the results of the fitting with the new hydrogen model spectra are substantially different from
 the model spectra computed without the Compton effect, as can be seen in Fig.\,\ref{fig8}. 
The new confidence regions are much more compact and are shifted toward lower 
masses and radii for the same assumed distances. 
This means that we would have to assume an even larger distance to HESS\,J1731$-$347  in order to explain the observed spectrum 
with the hydrogen model and assuming canonical NS mass and radius. 
The distances larger than 10\,kpc would, however, imply an unrealistically high TeV luminosity of the SNR \citep{Klochkov.etal:15}. 
 A morphological comparison of the photoelectric absorption in X-rays and the molecular gas density patterns led to a lower
  limit on the distance of $d=$\,3.2\,kpc, placing the remnant either in the Scutum-Crux Galactic spiral arm at a distance of 
  $\sim$3\,kpc or at the Norma-Cygnus arm at $\sim$4.5\,kpc \citep{HESS:2011}. 
A distance of 3.2\,kpc has been adopted in \citet{Cui:etal:16} and \citet{Doroshenko:etal:16}.
The best fit of the observed X-ray spectrum with the new set of the hydrogen model spectra with an assumed 
distance of 10\,kpc is shown in Fig.\,\ref{fig9}. 
The null-hypothesis probability of $P\sim 0.0002$ and  strong residual features indicate an unacceptable fit.
We note that the fit quality becomes even worse with increased distance.  
 
\begin{figure}
\centering
\includegraphics[angle=0,width=\columnwidth]{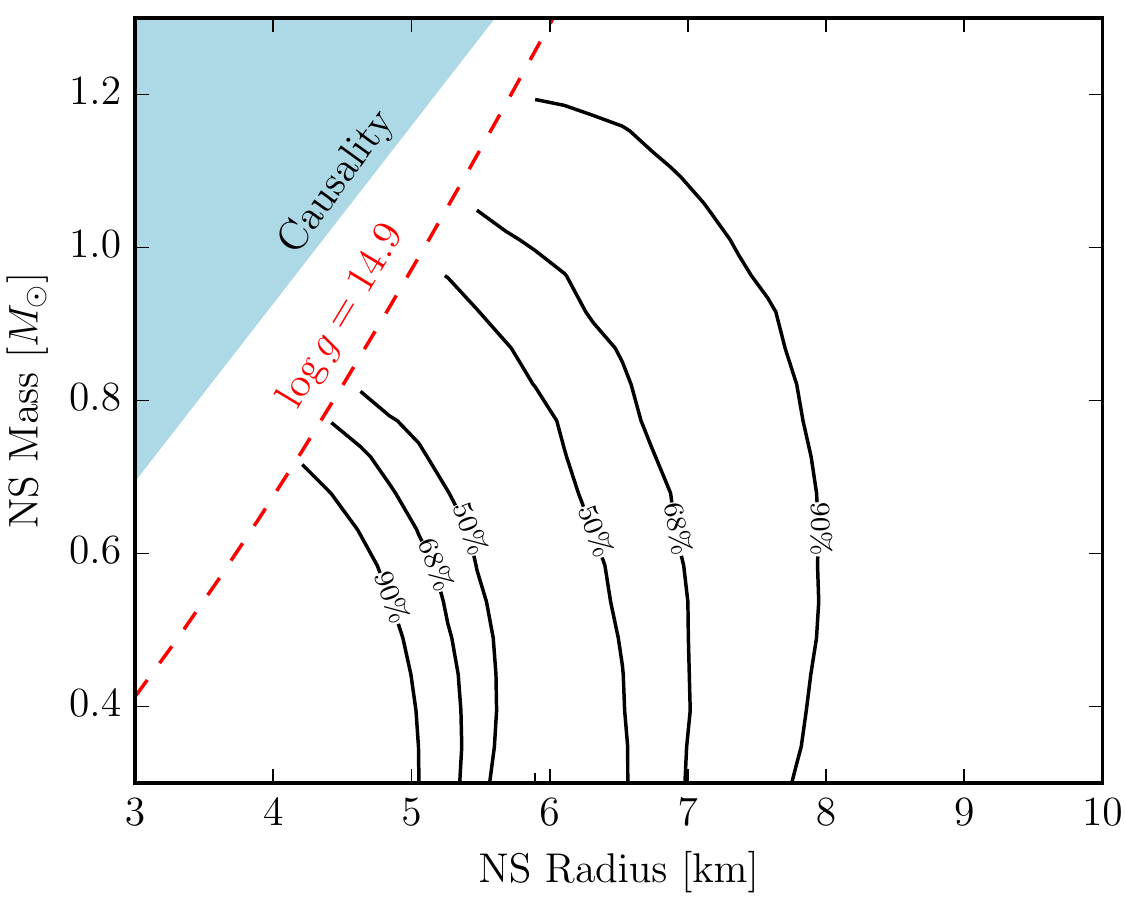}
\caption{\label{fig10}
Confidence contours on the mass--radius plane for the CCO in HESS\,J1731$-$347 (similar to Fig.\,\ref{figRMC})
using two emission components (the hotspots and the remaining stellar surface) 
of the pure hydrogen atmosphere models with Compton scattering. 
The assumed distance is 3.2\,kpc. 
}
\end{figure}

\begin{figure}
\centering
\includegraphics[angle=0,width=\columnwidth]{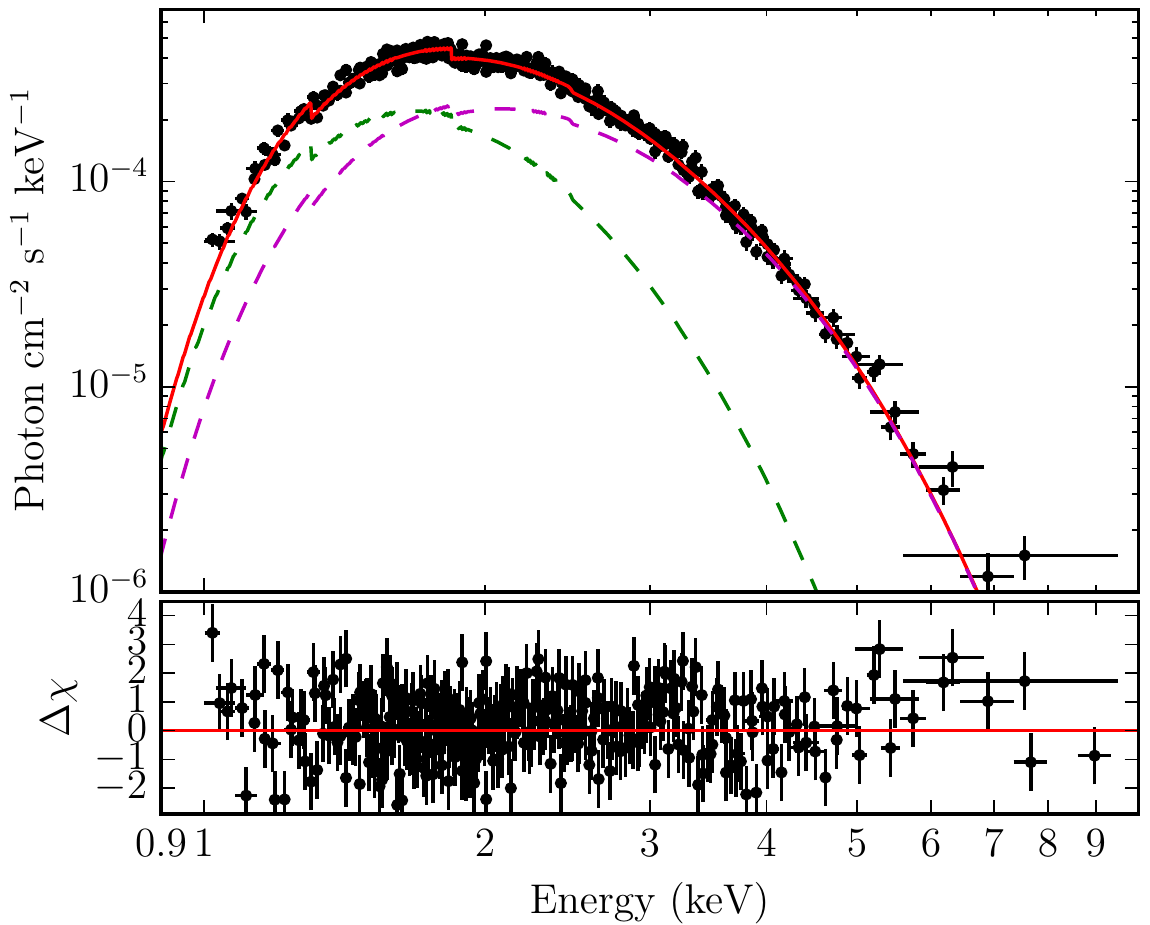}
\caption{\label{fig11}
Fit of the observed spectrum of the CCO in HESS\,J1731$-$347 using
two components (the hotspots and the remaining stellar surface) of the pure hydrogen atmosphere model  
spectra computed with Compton scattering. 
The assumed distance is 3.2\,kpc. 
The NS mass and radius are fixed to $M = 1.5 M_\odot$ and $R=12$\,km. 
The reduced $\chi^2$ is 1.08 (893 d.o.f.)  corresponding to the null-hypothesis probability of $P\simeq 0.04$.
The contribution of the cold and hot components are indicated with the dashed green and magenta curves, respectively.
}
\end{figure}

The existing upper limit on the $PF$ of $\sim$8\% evaluated in \citet{Klochkov.etal:15} does not allow us to formally
 exclude hotspots on the NS surface. 
To investigate the possibility of such hotspots, we fit a two-component model from our new grid 
of hydrogen model spectra to  the
observed spectrum. 
The model now consists of two model spectra of different temperatures with a free ratio of their emitting areas: 
\be \label{eq:twocomp}
  f'_{E} = \frac{R^2}{d^2(1+z)}\,\left(\delta\ F_{E'}(T_{1})+(1-\delta)\ F_{E'}(T_{2})\right) . 
\ee 
The emitted $E'$ and the observed $E$ photon energies   are connected by the relation $E'=E(1+z)$, 
where $z=(1-R_{\rm  S}/R)^{-1/2}-1$ and  $R_{\rm S} = 2GM/c^{2}$ is the Schwarzschild radius.
As expected, the resulting confidence regions become larger in this case because of the additional free parameters (see Fig.\,\ref{fig10}).
The quality of the two-component fit  (see Fig.\,\ref{fig11}) is better than that with one component with $\chi^2_{\rm red}$=1.08. 
However,  the NS canonical mass and radius $M=1.5 M_\odot$ and $R=12$\,km lie outside the
 90\% confidence regions for the assumed distance of 3.2\,kpc. 

The intrinsic effective temperatures of the components are $T_1 =4.55$\,MK and $T_2=2.04$\,MK, the best-fit
 fractional emitting area 
of the hot component is $\delta = 0.023\pm 0.003$  at the best-fit value of the interstellar hydrogen column density
 $N_{\rm H} = 2.13 \times 10^{22}$\,cm$^{-2}$. 
The hot component flux contribution to the total spectrum is about 40\%. 
We thus expect strong pulsation of the flux if such a two-component model is correct. 
We cannot exclude, however, some fine-tuned geometries that could lead to the low observed $PF$.  
It is possible, for example, that the rotation axis is almost co-aligned with the line of sight or that the hotspots are
 located close to the NS rotation axis. 
In the following, we investigate these possibilities.   

\subsection{Model light curves and NS geometry limitations} 

\begin{figure}
\centering
\includegraphics[angle=0,width=\columnwidth]{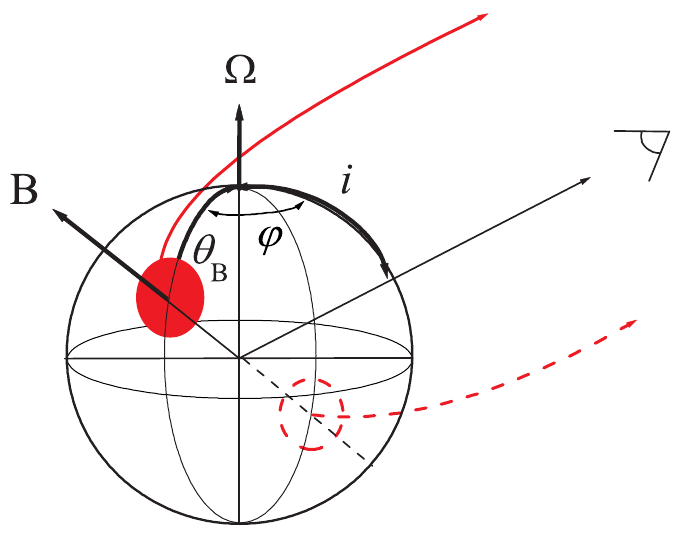}
\caption{\label{fig12}
Geometry of the model.  
}
\end{figure}

Here we model light curves of a slowly rotating NS with two identical uniform symmetrically located hotspots 
 using the approach described in \citet{Suleimanov.etal:10}. 
The  main model parameters are the NS mass $M$ and radius $R$, which determine the gravitational redshift $z$
 and the gravity $g=GMR^{-2}(1+z)$ on the stellar surface.  
In addition to these parameters, the model includes three geometrical parameters (see Fig.\,\ref{fig12}): the angle  between
 the rotation axis and the line of sight $i$, the angle between the rotation axis and the nearest hotspot center
  $\theta_{\rm B}$, and the angular radius of the spot $\theta_{\rm sp}$. 
The remaining two parameters
are the effective temperature of the spot
$T_{\rm sp}$ and the effective temperature of the remaining stellar
surface $T_{\rm c}$. 

We take the same NS basic parameters $M=1.5\, M_\odot$ and $R=12$\,km
for our investigations as we used for the two-component fit
described above. The corresponding values of the gravitational redshift
and the surface gravity are $z=0.26$ and $\log g =14.24$.  
We take the model spectra with the parameters closest to those obtained from
the two-component fits, 
$T_{\rm sp} = 4.55$\,MK and $T_{\rm c}=2.05$\,MK at $\log g =14.3$.

The model spectrum at the fixed $\theta_{\rm B}$, $i$ and the
phase angle $\varphi$  is computed according to the following formulae \citep{B02,PB06}
\be
       f_{E} (\varphi) = 
       \frac{R^2}{d^2(1+z)^{3}} \int_0^{2\uppi}\! \! \!  \!  \! {\rm d}\phi  \!  \!  \int_0^\uppi \! \!  \!  \!  I_{E'}(\alpha)
       \cos\alpha \sin\gamma\,{\rm d}\gamma,  
\ee
where $\phi$ and $\gamma$ are the azimuthal angle and colatitude in the spherical coordinate system with
 the $z$-axis along the NS rotation axis.
The distance is assumed to be 3.2 kpc. 
We assume that the NS rotates slowly. The code we used  is described in detail by \citet{Suleimanov.etal:10}. 
A photon emitted at an angle $\alpha$ to the local surface normal arrives at a distant observer at an angle $\psi$ to the local surface
 normal as a result of the light bending in the NS gravitational field. 
We use an approximate relation suggested by \citet{B02} to connect these angles
\be
 \cos\alpha \approx \frac{R_{\rm S}}{R}+ \left(1 - \frac{R_{\rm S}}{R}  \right) \cos\psi ,
\ee
where 
\be
 \cos\psi = \cos i\, \cos\gamma + \sin i\, \sin\gamma\,\cos\phi.
\ee
Only the NS area elements  with a local $\cos\alpha \ge 0$ contribute to the observed spectrum.  
The model spectrum $I_{E'}(\alpha)$ is computed using the tabulated model spectra with $\mu=\cos\alpha$ 
taking the hot model atmosphere with $T_{\rm eff} =T_{\rm sp}$ when the point with coordinates $\gamma$ and $\phi$
 is inside the hotspots, and using the cold model spectrum with $T_{\rm eff} =T_{\rm c}$  in the opposite case. 
The final model spectrum is corrected for interstellar absorption using the observed $N_{\rm H} = 2.13 \times 10^{22}$\,cm$^{-2}$. 
The corresponding absorption cross-sections were computed using the {\sc xspec} model {\sc xszvab} 
with a solar abundance of chemical elements. 

\begin{figure}
\centering
\includegraphics[angle=0,width=\columnwidth]{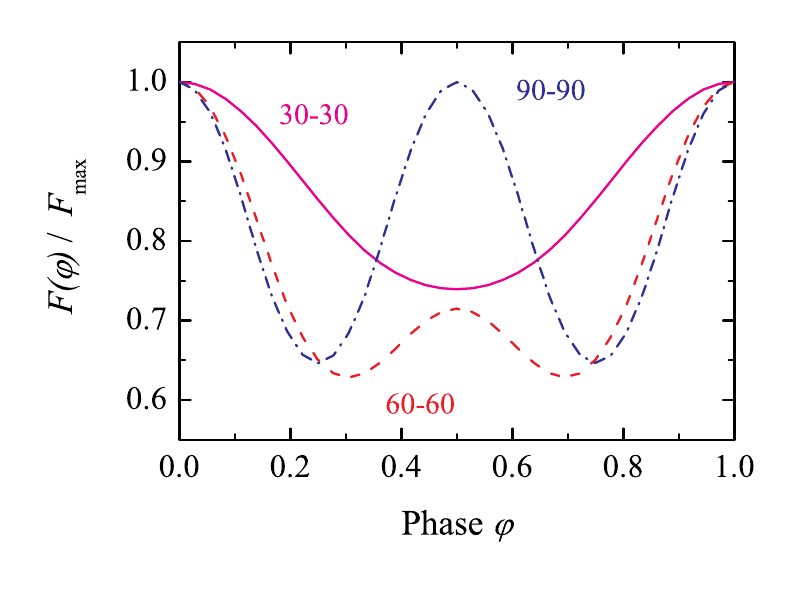}
\caption{\label{fig13}
Examples of the normalized light curves for the adopted NS and
hotspot parameters (see text). 
The curves are computed for angle pairs $i=30\degr$, $\theta_{\rm B}=30\degr$ (solid curve), 
$i=60\degr$, $\theta_{\rm B}=60\degr$ (dashed curve), and $i=90\degr$, $\theta_{\rm B}=90\degr$ (dash-dotted curve).
The spot angular radii $\theta_{\rm sp}$ and the $PF$ values are
11\degr and 15\%,  
12.3\degr and 23\%, and 11.5\degr and 21.4\%, respectively. 
}
\end{figure}

We did not fix the hotspot radius $\theta_{\rm sp,}$ but determined
it 
by fitting the phase-averaged spectrum to the reference spectrum
in the energy band 1--10 keV 
using the dichotomy method. 
The reference spectrum in the energy band 1--10 keV was computed
using Eq.\,(\ref{eq:twocomp})  with $T_1= T_{\rm sp}$ and $T_2=T_{\rm c}$
and compared to the phase-averaged spectra of the NS model for different spot radii. 
The interstellar absorption for the reference spectrum was also taken into account in the same way as for the model spectra.
We took the current spot radius to be the correct spot radius $\theta_{\rm sp}$ at given $i$ and $\theta_{\rm B}$ 
if the relative average deviation of the current phase-averaged spectrum from the reference spectrum was lower than 1\%. 
For the considered NS model, the actual spot radius changes from 10\degr\, to 16\degr\, for different values 
of  $i$ and $\theta_{\rm  B}$. 
Examples of the computed normalized light curves in the 1$-$10\,keV energy band are shown in Fig.\,\ref{fig13}.

\begin{figure}
\centering
\includegraphics[angle=0,width=\columnwidth]{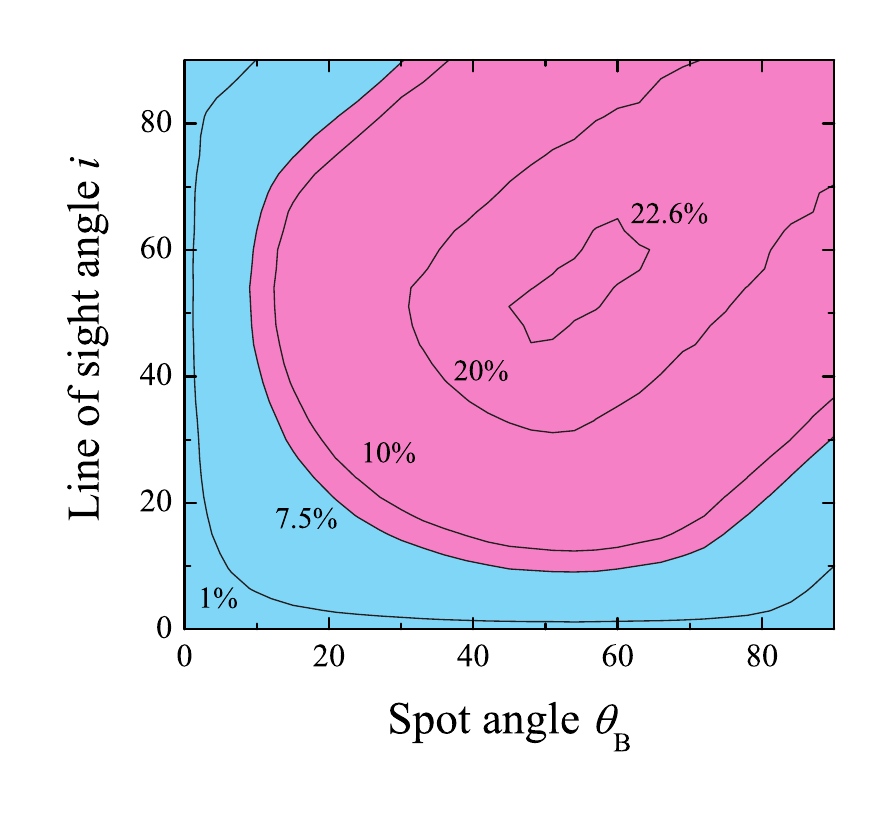}
\caption{\label{fig14}
Contours of constant  $PF$ in the $\theta_{\rm B} - i$ plane.  
The permitted region with $PF<7.5\%$ is shown in blue. 
The total probability for the angles to fall in the permitted region is about 8.2\% 
assuming a random observer inclination as well as a random angle between magnetic field and rotation axis. 
}
\end{figure}

Using the described method, we computed an extended set of light curves for the given NS parameters for all the possible 
$i$ and $\theta_{\rm B}$ and determined the corresponding $PF$ in the same energy band, 1--10 keV. 
The obtained $PF$ map in the $\theta_{\rm B} - i$ plane is shown in Fig.\,\ref{fig14}. 
Contours  of equal $PF$ are indicated. 
The maximum possible $PF$ is about 25\%. 
The region forbidden by the  observations, with $PF > 7.5\%$, is shown in pink.

These data allow us to evaluate the probability $p$ that the absence of observed pulsations from the CCO 
in HESS\,J1731$-$347 is due to an unfavorable combination of $i$ and $\theta_{\rm B}$ angles. 
Assuming  random observer inclination as well as random magnetic field inclination to the rotation axis, we have
\be
  p = \frac{\int_0^{\uppi/2}\int_0^{\uppi/2}\,S\,\sin i\,\sin\theta\, 
  {\rm d}\theta\,{\rm d}i}{\int_0^{\uppi/2}\int_0^{\uppi/2}\,\sin i\,\sin\theta\, {\rm d}\theta\,{\rm d}i},
\ee  
where $S =1$ if $PF \le 7.5\%$ for the given $i$ and $\theta_{\rm B}$ values and $S=0$ else. 
As a result, we obtain a relatively low probability of $p \approx 8.2\%$. 
It is, however, clearly not sufficiently low to exclude the possibility of the special geometry.

\section{Summary}
 
We computed  new model grids for pure carbon and pure hydrogen  NS atmospheres that account for Compton scattering using 
exact redistribution functions as well as using the Kompaneets equation. 
To model the radiation from a non-uniformly emitting NS, we also used the angular dependence of the emerging radiation. 
The new carbon atmosphere models differ only slightly from previous models, where electron scattering 
 was considered to be coherent. 
The new hydrogen atmosphere model spectra, however, differ substantially at effective temperatures $T_{\rm eff}>\,2$\,MK.  
We note that the angular distribution of the emerging spectra of the carbon model atmospheres at 
 photon energies $1-5$ keV is peaked 
more strongly along the surface normal than the  electron-scattering  dominated atmosphere.  
 
We have used the new sets of model atmospheres to fit the observed spectrum of the NS in
 the center of the supernova remnant HESS\,J1731$-$347. 
A fit with our new carbon spectra is almost identical to the fit published previously by \citet{Klochkov.etal:15}. 
The fits with the new hydrogen atmosphere spectra deviate substantially from those presented in \citet{Klochkov.etal:15}, 
resulting in  much more compact confidence regions in the $M-R$ plane, which are shifted
 to smaller NS radii than those obtained with the old models.
Nevertheless, the formal quality of the fits with the hydrogen models is poorer than that of the carbon atmosphere spectra. 
Moreover, an uncomfortably large distance to the source, $> 10$\,kpc, is required to be consistent with the canonical NS radius. 
Such a large distance contradicts other observational properties of the supernova remnant.

We have investigated a hypothesis that assumes that the observed radiation is the sum of two spectral components, from the 
hotspots and from the remaining stellar surface.
With the sum of two hydrogen atmosphere model spectra, we find an acceptable fit, which yields an unrealistically low
 NS mass and a too small radius for the adopted distance of 3.2\,kpc,
however.  
Fixing  $M=1.5\, M_\odot$, $R=12$\,km, and $d=3.2$\,kpc still results in a formally acceptable fit.
The effective temperature of the hot component is 4.55\,MK, its apparent fractional area is about 2.3\% of the total stellar 
surface, and  the angular radius of the hotspots is $\approx 12.4\degr$. 
The cold component corresponding to the remaining stellar surface has an effective temperature of 2.04\,MK. 

Using these values, we have evaluated the probability that we do not observe pulsations from the CCO in HESS\,J1731$-$347 
as a result of unfavorable geometry, namely due to a low inclination of the NS spin axis to the line 
of sight or due to a small angular distance 
of the hotspots from the NS rotation poles. 
To verify this, we modeled light curves of a slowly spinning NS.
We find that the probability of such a geometry by chance is relatively low, $\sim$8.2\%. 
We note that there exists another non-pulsating CCO in the supernova remnant Cas~A, 
whose spectrum was successfully described using 
a carbon atmosphere model, and which has a similar upper limit on the $PF$.
Because the probability of an unfavorable  geometry for Cas~A is comparable to the probability found here, 
the joint probability for these two NSs to have an unfavorable geometry
would be below 1\%.

We note here that we used non-magnetic  atmosphere models in our investigations. Hotspots on the NS surface can arise as a
result of a strong magnetic field. 
In this case,  using magnetized model atmospheres would be more appropriate. 
However, the magnetic field strength is not known. 
Computing magnetic atmospheres also requires much more extended studies, which are beyond the scope of the present work and 
will be considered in the future.

\begin{acknowledgements} 
The work was supported by the Deutsche Forschungsgemeinschaft (DFG) grant WE 1312/48-1
 and  Russian Science Foundation grant 14-12-01287 (VFS), 
the Magnus Ehrnrooth Foundation,  
the Foundations' Professor Pool, the Finnish Cultural Foundation, the Academy of Finland grant 268740 (JP). 
\end{acknowledgements}

\bibliographystyle{aa}
\bibliography{cmpCH}

\end{document}